\documentclass[reprint,preprintnumbers,nofootinbib,amsmath,amssymb,aps]{revtex4-2}
\usepackage{graphicx}
\usepackage{dcolumn}
\usepackage{bm}
\usepackage{amsmath}
\usepackage{caption}
\newcommand{\gsim}{\;\rlap{\lower 3.5 pt \hbox{$\mathchar \sim$}} \raise 1pt
\hbox {$>$}\;}
\newcommand{\lsim}{\;\rlap{\lower 3.5 pt \hbox{$\mathchar \sim$}} \raise 1pt
\hbox {$<$}\;}

\begin{document}
\allowdisplaybreaks

\preprint{ALBERTA-THY-3-22}

\title{On Evaluation of Nonfactorizable Corrections to Higgs Boson Production via Vector Boson Fusion}
\author{Logan Gates}
\altaffiliation{}
\email{lgates@ualberta.ca}
\affiliation{Department of Physics, University of Alberta, Edmonton, Alberta T6G 2J1, Canada}


\begin{abstract}
We present the fully analytical result for the
next-to-next-to-leading nonfactorizable QCD corrections to
Higgs boson production via vector boson fusion computed to the
leading power in the ratio of the jet
transverse momentum to the partonic center-of-mass energy.
\end{abstract}

\maketitle

The experimental observation of the Higgs boson in 2012 \cite{Higgs_discovery1,Higgs_discovery2} commenced the next
phase of the Large Hadron Collider (LHC) in undergoing
precision studies to measure the properties of the Higgs boson
and that will soon reach the milestone of the high-luminosity
programme with an objective to precisely determine all of its
fundamental parameters \cite{MICCO2020100045}. Vector boson
fusion (VBF) is the second most dominant production mode of the
Higgs boson with high sensitivity to the electroweak Higgs
couplings and a distinguishable final state characterized by
the presence of two energetic jets in the forward and backward
region of the detector
\cite{PhysRevD.92.032008,PhysRevD.98.052003,higgs_prop1,higgs_prop2}.

Evaluation of the high-order radiative corrections in the
strong coupling constant $\alpha_s$ is mandatory to get an
accurate theoretical description of the process and to confront
the Standard Model predictions with the current and future
experimental data. For a long time, the main obstacle in
advancing the perturbative QCD analysis of the Higgs boson VBF
production beyond the next-to-leading order (NLO) was the
so-called nonfactorizable contribution with gluon exchange
between the incoming quark lines. Due to color conservation,
this contribution starts at the next-to-next-to-leading order
(NNLO) and has been previously neglected due to the formal
$1/N_c^2$ suppression in the limit of large number of colors
$N_c\to \infty$ in comparison to its factorizable counterpart,
known through $\alpha_s^3$ for Higgs boson production
\cite{color_supp,NNLO,N3LO,PhysRevD.68.073005,PhysRevLett.115.082002}. 
Though the technical difficulty of the exact calculation in the
case of the nonfactorizable NNLO corrections is still beyond
the limit of existing computational methods, they have been
recently evaluated for single Higgs production at the leading
power of the expansion in the ratio of the jet transverse
momentum to the partonic center-of-mass energy
$p_{j,\perp}/\sqrt{\hat{s}}$ \cite{Penin}, which is small in
the VBF kinematics. The resulting corrections are closely
related to the incomplete cancellation of the Glauber phase and
receive a $\pi^2$ enhancement characteristic to a scattering
phase that largely compensates the color suppression. The
corrections to the differential cross section may exceed 10\%
in the region $m^2_V<p^2_{j,\perp}\ll\hat{s}$, where $m_V =
M_{Z,W}$ is the vector boson mass. Thus, they have to be taken
into account to reach a subpercent precision of the theoretical
predictions in the whole experimentally relevant phase space.

However, Ref.~\cite{Penin} only provides the result for
the NNLO corrections to the cross section in the form of an
integral representation. In the present paper, we complete this
analysis  and present the fully analytic expression 
for the corrections.

We adopt the following notation for single Higgs VBF production
in almost forward scattering of two highly energetic quarks
\begin{equation*}
	Q_1(p_1)+Q_2(p_2) \rightarrow Q_1(p_3)+Q_2(p_4)+H(p_5)\, . 	
\end{equation*}
The incoming quark momenta are chosen to have only the
light-cone components $p_1^\mu =(0,p_1^-,0)$ and $p_2^\mu
=(p_2^+,0,0)$ so that in the VBF kinematics the momentum
transfers read $q_1^\mu =p_1^\mu -p_3^\mu\approx(0,q_1^-,q_1)$
and $q_2^\mu =p_2^\mu -p_4^\mu\approx (q_2^+,0,q_2)$. For a
four-vector $k^\mu=(k^+,k^-,k)$ in the light-cone coordinates,
$k=(k^1,k^2)$ stands for its transverse components.

With the VBF cuts, the characteristic scales of the energy and
transversal momentum of the tagging jets are $\sqrt{\hat{s}}
\gsim 600~{\rm GeV}$ and $p_{j,\perp} \sim 100~{\rm GeV}$
\cite{PhysRevD.92.032008,PhysRevD.98.052003}, making the ratio
$p_{j,\perp}/\sqrt{\hat{s}}$ a good expansion parameter
\cite{Penin:2019xql}. The leading order of such an expansion
defines the eikonal approximation
\cite{eikonal1,eikonal2,eikonal3}. In this approximation, the
light-cone and transversal degrees of freedom decouple, and the
process can be described by an effective theory involving the
on-shell quarks propagating along the light cone and
``Glauber'' gauge bosons propagating in the two-dimensional
transversal space.

\begin{figure}[t]
	\begin{center}
		\begin{tabular}{ccc}
			\hspace*{-8mm}\includegraphics[width=4.6cm]{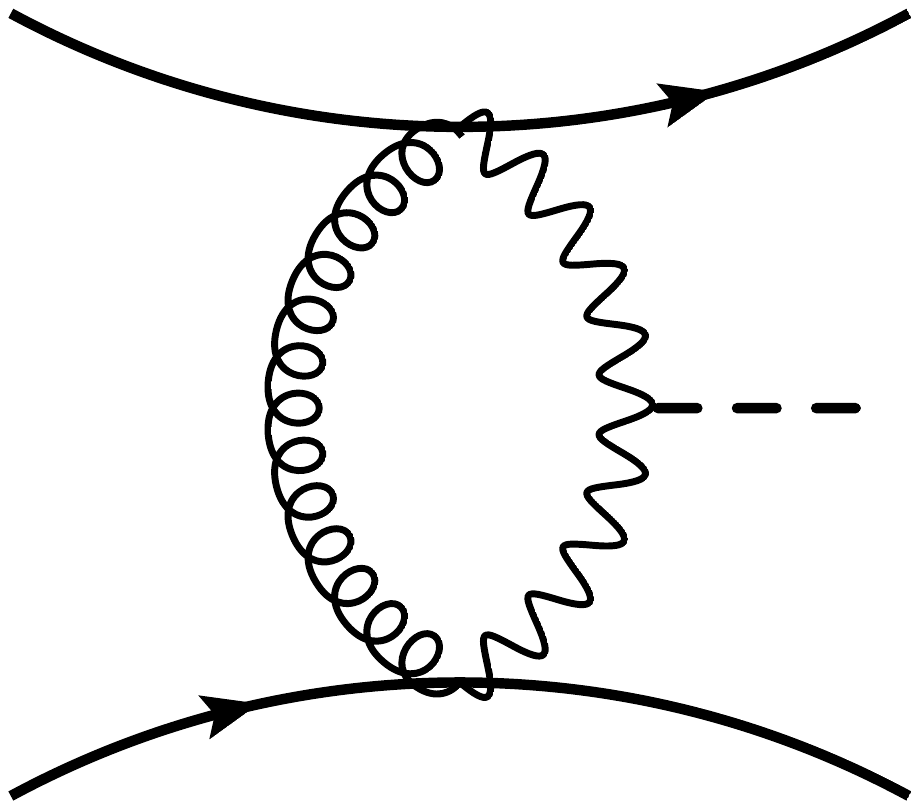}&
			\hspace*{-8mm}\includegraphics[width=4.6cm]{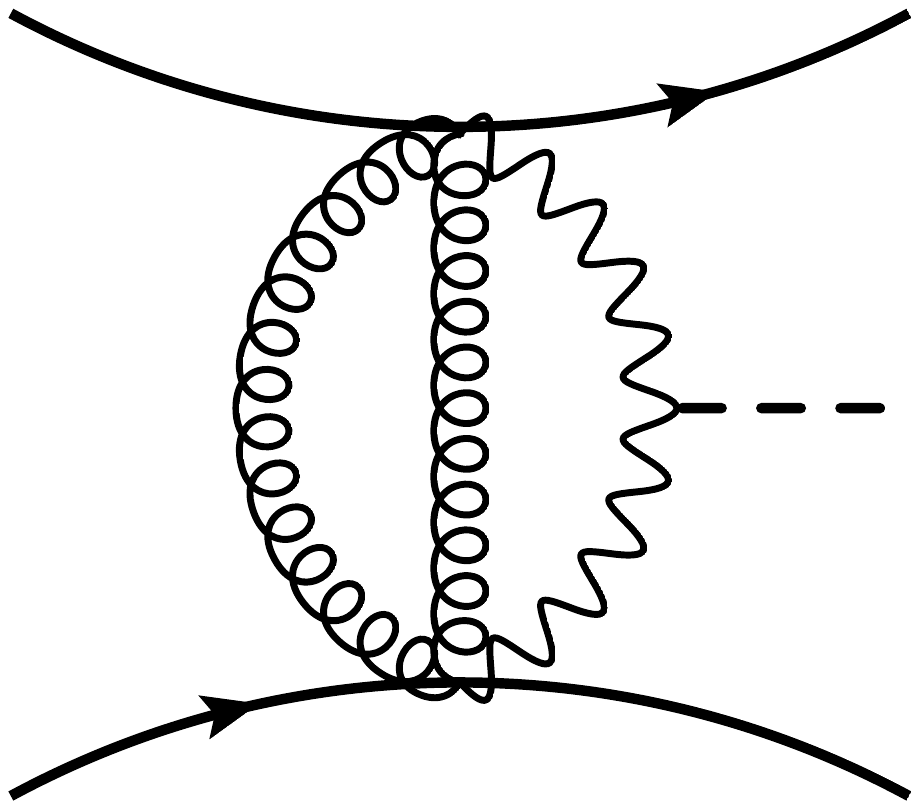}
			\\[-22mm]
			\hspace*{-5mm}(a)&\hspace*{-5mm}(b)\\
		\end{tabular}
	\end{center}
	\caption{\label{fig:effdiag}  (a) One- and (b) two-loop 	
    contributions to the Glauber phase for the Higgs boson 	
    production. The loopy, wavy, and dashed lines correspond to the
    gluon, vector boson, and Higgs boson, respectively.}
\end{figure}

In the leading eikonal approximation, the NNLO nonfactorizable
correction to the differential cross section reads \cite{Penin}
\begin{eqnarray}
	{\rm d}\sigma^{\rm NNLO}_{\rm nf}&=&
	\left({N_c^2-1\over 4N_c^2}\right)\alpha_s^2\,
	\chi_{\rm nf}\,
	{\rm d}\sigma^{\rm LO}\,.
	\label{eq::sigma}
\end{eqnarray}
Here ${\rm d} \sigma^{\rm LO}$ is the leading-order cross section for VBF and
\begin{equation}
	\chi_{\rm nf} =\left[\chi^{(1)}\right]^2 - \chi^{(2)}\,,
	\label{eq::chinf}
\end{equation}
where $\chi^{(1)}$ and $\chi^{(2)}$ are the one- and two-loop
contributions to the Glauber phase acquired by the scattering
quarks. They are determined by the two-dimensional effective
theory Feynman diagrams in Fig.~\ref{fig:effdiag}~(a,b) which give
\begin{eqnarray}
	\chi^{(1)}&=&\frac{1}{\pi}\int\frac{d^2k}{k^2+\lambda^2} \nonumber \\
	&& \times \frac{\Delta_1}{(k-q_1)^2+m_V^2}\frac{\Delta_2}{(k+q_2)^2+m_V^2}\,,  \label{eq::chi1int}\\ \chi^{(2)}&=&\frac{1}{\pi^2}\int\frac{d^2k_1}{k_1^2+\lambda^2}\frac{d^2k_2}{k_2^2+\lambda^2} \nonumber \\
	&& \times \frac{\Delta_1}{(k_{12}-q_1)^2+m_V^2}\frac{\Delta_2}{(k_{12}+q_2)^2+m_V^2}\,,
	\label{eq::chi2int}
\end{eqnarray}
where $k_{ij}= k_i +k_j$, $\Delta_i = q_i^2 +m_V^2$, and $\lambda$
is the fictitious gluon mass that is introduced to regulate the
infrared divergence. The singular dependence of the one- and
two-loop contributions on the infrared cutoff can be separated
as follows
\begin{eqnarray}
	\chi^{(1)}&=&- \ln \left(\frac{\lambda^2}{m_V^2}\right) +f^{(1)}\,,
  \label{eq::chi1dec}\\
	\chi^{(2)}&=& \ln^2 \left(\frac{\lambda^2}{m_V^2}\right) -2 \ln\left(\frac{\lambda^2}{m_V^2}\right) f^{(1)} + f^{(2)}\,.
	\label{eq::chi2dec}
\end{eqnarray}
The functions $f^{(i)}$ do not depend on $\lambda$, and the
singular terms cancel out in the expression for $\chi_{\rm
nf}=\left[f^{(1)}\right]^2-f^{(2)}$. The three-point Feynman
integrals in two dimensions can be reduced to the linear
combinations of the two-point integrals \cite{loop_method}.
Hence, Eqs.~(\ref{eq::chi1int},\ref{eq::chi2int}) can be
integrated through two loops in terms of dilogarithmic
functions. For the infrared finite part of the one-loop
contribution, we find
\begin{eqnarray}
	f^{(1)}&=&\frac{\Delta_1\Delta_2}{\Delta_{T}}\left[2\frac{q_1^2\left( -q_1^2+q_2^2+s \right)+m_V^2\left( q_1^2-q_2^2+s\right)}{\Delta_1} \right. \nonumber\\
	&& \times \ln\left(\frac{\Delta_1}{m_V^2} \right) + \frac{s\left(q_1^2+q_2^2-s-2m_V^2\right)}{\gamma} \ln\left( \frac{\gamma + s}{\gamma -s} \right) \nonumber\\
	&&+ (q_1\leftrightarrow q_2,~\Delta_1\leftrightarrow \Delta_2)\Bigg]\,,
	\label{eq::f1}
\end{eqnarray}
with $\gamma=\sqrt{s\left(s+4m_V^2\right)}$, $s=(q_1+q_2)^2$
is the Higgs bosons transverse momentum ($p_5=-q_1-q_2$)
squared, and
\begin{eqnarray}
	\Delta_{T}&=&2\left[ q_1^2q_2^2s+ m_V^4s+m_V^2 \left( 2q_1^2q_2^2 + s\left(q_1^2+q_2^2\right) \right.\right.\nonumber\\
	&& \left.\left. -q_1^4-q_2^4 \right)  \right]\,.
\end{eqnarray}
This result agrees with the expression obtained in
Ref.~\cite{Oxford}.  The finite part of the two-loop correction
to the Glauber phase reads
\begin{widetext}
\begin{eqnarray}
f^{(2)}&=&-4f^{(1)} + \frac{\pi^2}{3} - 3 +
\frac{3}{\Delta_1-\Delta_2} \left[ \Delta_1\left(\ln\left(\frac{\Delta_2}{m_V^2}\right)+1\right)^2 - \Delta_2\left(\ln\left(\frac{\Delta_1}{m_V^2}\right)+1\right)^2 \right] \nonumber \\
&& + \frac{4 \Delta_1 \Delta_2}{\mu \gamma} \left[ \gamma^2 - \left(\Delta_1+\Delta_2\right)s \right] \ln\left(\frac{\gamma+s}{\gamma-s}\right)+\frac{2\Delta_1 \Delta_2}{\mu \left(\Delta_1-\Delta_2\right)}\left[ \left(\Delta_1-\Delta_2\right)^2 - \left( \Delta_1+\Delta_2 \right)s \right] \nonumber\\
&& \times \left[ \left( \ln\left(\frac{\Delta_1}{m_V^2}\right) +1 \right)^2 - \left( \ln\left(\frac{\Delta_2}{m_V^2} \right) +1 \right)^2 \right]-\frac{4s\Delta_1 \Delta_2}{\Delta_1 - \Delta_2} \Bigg[\frac{1}{\gamma\left(\Delta_1-\Delta_2\right)+\left(\Delta_1+\Delta_2\right)s} \nonumber\\
&& \times \left(\text{Li}_2\left(\frac{2s\Delta_1}{\gamma\left(\Delta_1-\Delta_2\right)+s\left(\Delta_1+\Delta_2\right)}\right)-\text{Li}_2\left(\frac{2s\Delta_2}{\gamma\left(\Delta_1-\Delta_2\right)+s\left(\Delta_1+\Delta_2\right)}\right)+  \ln\left(\frac{\Delta_1}{m_V^2}\right) \right. \nonumber \\
&& \times \left.  \ln\left(\frac{\left(\Delta_1-\Delta_2\right)(\gamma-s)}{\gamma\left(\Delta_1-\Delta_2\right) +\left(\Delta_1+\Delta_2\right)s}\right)-\ln\left(\frac{\Delta_2}{m_V^2}\right)\ln\left(\frac{\left(\Delta_1-\Delta_2\right)(\gamma+s)}{\gamma\left(\Delta_1-\Delta_2\right)+\left(\Delta_1+\Delta_2\right)s}\right)\right)  \nonumber \\
&& -(s \leftrightarrow -s) \Bigg] -\frac{\Delta_1 \Delta_2}{\gamma \mu} \Bigg[
\left(\gamma\left(\Delta_1-\Delta_2\right)-\left(\Delta_1+\Delta_2\right)s\right)\left(\ln\left(\frac{\Delta_1}{m_V^2}\right)\ln\left(\frac{\left(\Delta_1-\Delta_2\right)(\gamma-s)}{\gamma\left(\Delta_1-\Delta_2\right)+\left(\Delta_1+\Delta_2\right)s}\right)\right. \nonumber\\
&& -\ln\left(\frac{\Delta_2}{m_V^2}\right) \ln\left(\frac{\left(\Delta_1-\Delta_2\right)(\gamma+s)}{\gamma\left(\Delta_1-\Delta_2\right)+\left(\Delta_1+\Delta_2\right)s}\right) + \text{Li}_2\left(\frac{2s\Delta_1}{\gamma\left(\Delta_1-\Delta_2\right)+s\left(\Delta_1+\Delta_2\right)}\right) \nonumber \\
&& - \left.  \text{Li}_2\left(\frac{2s\Delta_2}{\gamma\left(\Delta_1-\Delta_2\right)+s\left(\Delta_1+\Delta_2\right)}\right)\right) + (s\leftrightarrow -s) \Bigg] -\frac{2 \Delta_1 \Delta_2}{\Delta_1-\Delta_2} \left[ \frac{\ln\left(\frac{\Delta_1}{m_V^2}\right)-\text{Li}_2\left(\frac{q_1^2}{\Delta_1}\right)}{\Delta_1} \right.\nonumber \\
&& \left.-\frac{\ln\left(\frac{\Delta_2}{m_V^2}\right)-\text{Li}_2\left(\frac{q_2^2}{\Delta_2}\right)}{\Delta_2} \right] -
\frac{\Delta_1 \Delta_2}{ \mu \left( \Delta_1 - \Delta_2 \right)} \Bigg\{ \Bigg[  \left(-\left(\Delta_1-\Delta_2\right)\zeta+\left(\Delta_1-\Delta_2\right)^2-\left(\Delta_1+\Delta_2\right)s\right) \nonumber \\
&& \times \left(\ln\left(\frac{\Delta_1}{m_V^2}\right)\ln\left(\frac{\Delta_1-\Delta_2+\zeta+s}{m_V^2}\right)-\ln\left(\frac{\Delta_2}{m_V^2}\right) \ln\left(\frac{\Delta_1-\Delta_2+\zeta-s}{m_V^2}\right)\right)+(\zeta \leftrightarrow -\zeta) \Bigg] \nonumber\\
&& +2 \left[\left(\Delta_1-\Delta_2\right)^2-\left(\Delta_1+\Delta_2\right)s\right]\Bigg[ \text{Li}_2\left(-\frac{2s\Delta_2}{\gamma\left(\Delta_1-\Delta_2\right)-s\left(\Delta_1+\Delta_2\right)}\right) \nonumber \\
&& + \text{Li}_2\left(\frac{2s\Delta_2}{\gamma\left(\Delta_1-\Delta_2\right)+s\left(\Delta_1+\Delta_2\right)}\right) +\ln\left(\frac{\Delta_2}{m_V^2}\right)\left(\ln\left(\frac{\left(\Delta_1-\Delta_2\right)(\gamma-s)}{\gamma\left(\Delta_1-\Delta_2\right)-\left(\Delta_1+\Delta_2\right)s}\right) \right. \nonumber \\
&& \left.  + \ln\left(\frac{\left(\Delta_1-\Delta_2\right)(\gamma+s)}{\gamma\left(\Delta_1-\Delta_2\right)+\left(\Delta_1+\Delta_2\right)s}\right)\right)  -(\Delta_1 \leftrightarrow \Delta_2) \Bigg] \nonumber \\
&& + \Bigg[ \left(-\left(\Delta_1-\Delta_2\right)\zeta-\left(\Delta_1-\Delta_2\right)^2+ \left(\Delta_1+\Delta_2\right)s\right) \Bigg(\text{Li}_2\left(\frac{-s-\zeta+\Delta_1-\Delta_2}{\gamma-\zeta+\Delta_1-\Delta_2}\right) \nonumber \\
&& -\text{Li}_2\left(\frac{s-\zeta+\Delta_1-\Delta_2}{\gamma-\zeta+\Delta_1-\Delta_2}\right)-\text{Li}_2\left(\frac{-s+\zeta-\Delta_1+\Delta_2}{\gamma+\zeta-\Delta_1+\Delta_2}\right)+\text{Li}_2\left(\frac{s+\zeta-\Delta_1+\Delta_2}{\gamma+\zeta-\Delta_1+\Delta_2}\right) \nonumber \\
&& -\text{Li}_2\left(\frac{\left(\Delta_1-\Delta_2\right)\left(-s-\zeta+\Delta_1-\Delta_2\right)}{\left(\Delta_1-\Delta_2\right)^2-\zeta\left(\Delta_1-\Delta_2\right)-s\left(\Delta_1+\Delta_2\right)}\right)+\text{Li}_2\left(\frac{\left(\Delta_1-\Delta_2\right)\left(s-\zeta+\Delta_1-\Delta_2\right)}{\left(\Delta_1-\Delta_2\right)^2-\zeta\left(\Delta_1-\Delta_2\right)-s\left(\Delta_1+\Delta_2\right)}\right) \nonumber \\
&&  +\ln\left(\frac{\Delta_1-\Delta_2-\zeta+s}{m_V^2}\right) \Bigg(-\ln\left(\frac{\gamma-s}{\gamma+\Delta_1-\Delta_2-\zeta}\right)- \ln\left(\frac{\gamma+s}{\gamma-\Delta_1+\Delta_2+\zeta}\right) \nonumber \\
&& +    \ln\left(-\frac{2\Delta_1s}{-\left(\Delta_1-\Delta_2\right)\zeta+\left(\Delta_1-\Delta_2\right)^2-\left(\Delta_1+\Delta_2\right)s}\right)\Bigg)-\ln\left(\frac{\Delta_1-\Delta_2-\zeta-s}{m_V^2}\right) \nonumber \\
&& \times \Bigg(-\ln\left(\frac{\gamma+s}{\gamma+\Delta_1-\Delta_2-\zeta}\right)-\ln\left(\frac{\gamma-s}{\gamma-\Delta_1+\Delta_2+\zeta}\right) \nonumber \\
&& + \ln\left(-\frac{2\Delta_2s}{-\left(\Delta_1-\Delta_2\right)\zeta+\left(\Delta_1-\Delta_2\right)^2-\left(\Delta_1+\Delta_2\right)s}\right)\Bigg)\Bigg)+ (\zeta \leftrightarrow - \zeta, ~\gamma \leftrightarrow -\gamma) \Bigg] \Bigg\} \,,
\end{eqnarray}
\end{widetext}
where
$\mu=m_V^2\left(\Delta_1-\Delta_2\right)^2-s\Delta_1\Delta_2$
and
$\zeta=\left(\left(q_1^2-q_2^2-s\right)^2-4q_2^2s\right)^{1\over
2}$. We have verified that numerically our result agrees with
the integral representation of the two-loop contribution given
in Ref.~\cite{Penin}.

To conclude, we have refined the previous analysis of the
nonfactorizable effects in Higgs boson VBF production. The NNLO
corrections have been obtained in fully analytic form  to the
leading power in $p_{j,\perp}/\sqrt{\hat{s}}$, {\it i.e.}, in
the eikonal approximation. Apart from the general bias
towards the analytic results in perturbative QCD calculations, such
a form would be preferable for the practical implementation
into the event generators and far more convenient for the study
of various asymptotic limits that facilitate the qualitative
understanding of the nonfactorizable phenomena.

\vspace{5mm}
\noindent
{\bf Acknowledgments.} I would like to thank Alexander Penin
for useful discussions and for valuable comments on the
manuscript. This research was supported in part by NSERC.

\bibliography{NF_Higgs_VBF_Ref}

\end{document}